# User software for the next generation

T. G. Worlton[1], A. Chatterjee[1], J. P. Hammonds[1], P. F. Peterson[1], D. J. Mikkelson[2], and R. L. Mikkelson[2]
[1]*Argonne National Laborator*y, *Argonne, IL 60439, USA*
[2]*University of Wisconsin-Stout, Menomonie, WI, 54751*

*Abstract*--New generations of neutron scattering sources and instrumentation are providing challenges in data handling for user software. Time-of-Flight instruments used at pulsed sources typically produce hundreds or thousands of channels of data for each detector segment. New instruments are being designed with thousands to hundreds of thousands of detector segments. High intensity neutron sources make possible parametric studies and texture studies which further increase data handling requirements. The Integrated Spectral Analysis Workbench (ISAW) software developed at Argonne handles large numbers of spectra simultaneously while providing operations to reduce, sort, combine and export the data. It includes viewers to inspect the data in detail in real time. ISAW uses existing software components and packages where feasible and takes advantage of the excellent support for user interface design and network communication in Java. The included scripting language simplifies repetitive operations for analyzing many files related to a given experiment. Recent additions to ISAW include a contour view, a time-slice table view, routines for finding and fitting peaks in data, and support for data from other facilities using the NeXus format. In this paper, I give an overview of features and planned improvements of ISAW. Details of some of the improvements are covered in other presentations at this conference.

## I. Introduction

New generations of neutron sources are producing large quantities of scattering data which must be handled efficiently in order for scientists to be able to process their data as rapidly as they collect it. Rapid response is also needed in guiding the experimenter in the length of collection time needed and in quickly observing interesting changes in data that need to be investigated further.

The large data sets being produced by new instruments arise from several factors:
- Linear position sensitive detectors (LPSD's)
- Multiple area detectors
- Large detector arrays
- Single crystal measurements
- Multiple measurements per sample for parametric studies
- Separate storage of detector data for texture measurements
- Separate storage of detector data for diagnostic purposes

The increasing usage of these neutron sources by non-experts also means that the software for collecting and reducing the neutron scattering data needs to be easy to use. We can no longer assume that every user who does an experiment will be an experienced user willing and able to travel to the source to do the experiment. Remote access to data and the neutron scattering instrument will become more important as the neutron scattering community expands.

Software for this next generation of neutron sources must have the following characteristics:
- Be able to read, correct, transform, and view very large sets of data.
- Be easy to use
- Allow remote access to data
- Run on common computing platforms (Windows, UNIX/Linux, Mac)
- Allow writing simple scripts to allow handling a large number of files and/or operations automatically
- Be freely available to users

At IPNS we have developed Integrated Spectral Analysis Workbench software (ISAW) which satisfies these requirements. ISAW is written in the Java language which is freely available from the Sun web site (http://java.sun.com/). In order to encourage user contributions, we also make the ISAW source code freely available through a link on the ipns web site (http://www.pns.anl.gov/computing/). ISAW has been tested on Windows systems, Linux systems, and Macintosh systems. It includes an automatic installation procedure that can run on any of the above platforms as long as they have Java 1.3.1 or later. ISAW can read NeXus files and n-column ASCII files as well as

IPNS run files so it will be useable at other pulsed neutron sources as long as the data files contain the needed information. The package also includes client and server software for remote access to data. Provisions are made for access to live data as well as remote file data. The software is flexible enough that different types of data such as regular detector data, LPSD data, area detector data, log file data, and pulse height data can be read and displayed by ISAW. We have successfully read, viewed and transformed LANSCE data from NeXus files as well as data from IPNS run files for all IPNS instruments.

In this paper we discuss ways to handle and visualize the large data sets being produced by new instrumentation at existing neutron sources. We show examples of large data sets from parametric studies, texture studies with diffractometers, and single crystal diffraction measurements.

We give some estimates of data set sizes, and then discuss what is needed to be able to handle those data sets. We also discuss changes needed in order to handle even larger data sets.

## *II. Next generation data sets*

The high cost of intense neutron sources has made it cost effective to provide many individual detectors or area detectors to observe as many of the scattered neutrons as possible (J. M. Carpenter, 1967) and has led to large increases in the number of detector elements (W. G. Williams, et al., 1998). Part of this increase in quantity of data comes from the use of time-of-flight methods which cause an increase in the amount of data to be handled per detector by breaking the data up into hundreds or thousands of time channels per detector element (Buras, et. al. 1963, 1965).

In some cases it is possible to focus data from large detector arrays as it is being collected either through proper placement of the detectors or through making real-time corrections to the measured time (J.M. Carpenter, 1967). However even when such focusing corrections are made, it is not always desirable to add the data together before examining it. There may be anisotropies in the data, multiple scattering effects, or problems with the data acquisition hardware that argue for keeping the data from each detector segment or pixel separate until the isotropy of the data can be verified.

IPNS started operation in 1980 and was the first generation of several proton accelerator based pulsed neutron sources. Many IPNS instruments contain over 100 detectors, and the computers of the 1980's were only able to handle the data by having the Data Acquisition System sum banks of detectors which had been electronically time-focused (Carpenter, et al. 1972). For diffractometers, the correction factor used to focus each spectrum is given by $L_i sin\theta_i / L_r sin\theta_r$, where "i" indexes each detector in the bank and "r" is the reference angle. The reference angle is chosen as the center of the bank of detectors (Jorgensen et al., 1989). More recent experiments have required postponing the summing of the data until the data can be examined for anisotropy caused by sample texture or to examine the data for instrumental problems. The number of detector elements has also increased several times as money for new banks of detectors has become available. Table 1 shows some current IPNS data set size estimates and data rate estimates and Table 2 shows estimates for similar proposed instruments at the Spallation Neutron Source (SNS).

Table 1 Dataset sizes and data rates of some IPNS Instruments.

| Instrument | No. of pixels / segments | No. of Time channels | Dataset size (KB) | Collection Time |
|---|---|---|---|---|
| SCD | 7200 | 120 | 3456 | 10 hr |
| SEPD | 160 | 5000 | 80 | ~ 1 hr |
| GPPD | 150 | 5000 | 3000 | ~ 1 hr |
| HRMECS | 2000 | 3000 | 24000 | 1 day |
| GLAD | 15000 | 5000 | 6000 | 5 hr |

Table 2 Dataset sizes and data rates of some proposed instruments at SNS (Hodges, 2002)

| Instrument | No. of pixels (Kpx) | No. of Time channels | Dataset size (MB) | Collection Time |
|---|---|---|---|---|
| Single Crystal Diffractometer | 5,000 | 85 | 1,700 | 1 hr |
| High Pressure Diffractometer | 100 | 2000 | 800 | 3 hr |
| Engineering Diffractometer | 80 | 750 | 240 | 1 min |
| ARCS chopper Spectrometer | 70 | 3000 | 840 | 30 min |
| Disordered Materials Diffractometer | 50 | 2000 | 400 | 1 hr |

In some cases the IPNS Dataset size estimates have been reduced because of detector grouping. The dataset sizes only include the histogram arrays without other data. The amount of memory required to analyze the data will be several times larger.

### III. ISAW and large datasets

The design of ISAW includes a number of features designed to efficiently handle and view large numbers of spectra. Sets of spectra with the same units for the x axis can be combined into data sets which can be viewed or operated on. As data sets get larger, and we try to handle more data sets at a time, the limiting factor is normally available system memory.

The "Image View" in ISAW is efficient for interactively viewing large numbers of spectra as shown in Fig. 1. The image view is based on software previously developed at IPNS (D. J. Mikkelson et al., 1995). In this view, each spectrum occupies one or more horizontal lines of the image display. If there are only a few spectra in the data set, there will be a band of lines having the same coloring. When there are more spectra than horizontal scan lines available in the display, scroll bars appear to the right of the image to allow scrolling to the spectra that do not fit on the display. When necessary, the image is also compressed horizontally to fit within the available window area. If the user desires to see all data points, this compression can be turned off and horizontal scrolling can be used instead. As shown in Fig. 1, dead detectors show up as black horizontal lines.

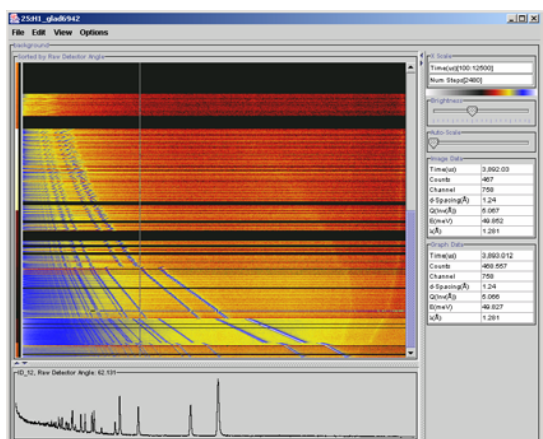

Figure 1. ISAW Image View of a powder calibrant in the Glass and Amorphous Systems Diffractometer (GLAD) at IPNS. The black horizontal lines represent dead detectors. The cross hairs show the position of the cursor over the image, and the right hand side contains display controls and the cursor readout area.

The graph at the bottom of the image view in Fig. 1 shows the spectrum being pointed at by the cursor. The right side of the image view in Fig. 1 contains a cursor readout which shows details of the data point at the cursor position. Both of these extra areas can be eliminated if the user desires to maximize the image area.

Another useful viewer for scattering instruments containing large detector banks is the "3D View" shown in Fig. 2. In this view, each detector element appears at the appropriate position in a 3D image and the brightness of the pixel corresponds to the number of counts for a given histogram bin. There are controls on the viewer which allow rotating the viewing position. There are also control buttons which allow stepping through the data along the x axis, in this case, time-of-flight. Displaying the total count by changing the number of bins for each detector element gives a nice picture of the detector layout for the instrument as well as the intensity of the scattering.

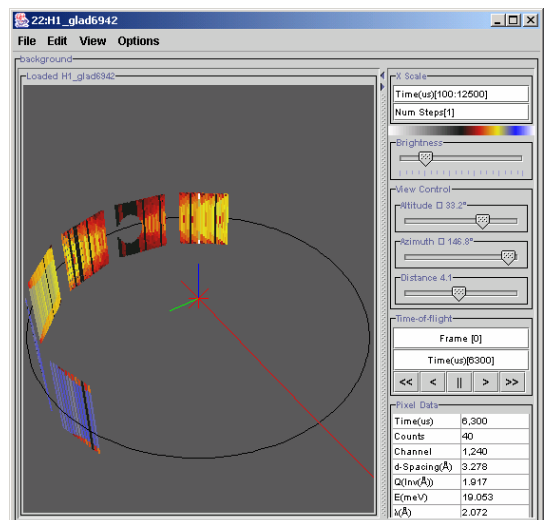

Figure 2. A 3D view of the same GLAD data shown in the image view of Fig. 1.

### IV. Parametric Studies with Diffractometers

ISAW was designed to simplify merging of data from a series of measurements to facilitate parametric studies. Such studies are not yet common at IPNS, but are expected to become common at next generation sources. One such study was done at the General Purpose Powder Diffractometer (GPPD) at IPNS. A sample of ceramic mixed oxide was studied as a function of Oxygen partial pressure using measurements lasting approximately seven minutes each. A script was written which allowed ISAW to read each of the 120 files, extract the 90 degree bank data, normalize the spectra, re-label the spectra, and merge them into a single data set. Since there was no other physical parameter in the data set, the label of each spectrum was set to be the run number and starting time of the measurement. Loading, merging, and viewing a single bank from 120 files takes approximately 30 seconds on a modern PC.

The result is shown in Fig. 3. The image view makes it obvious where the phase transition takes place, and cursor interaction with the plot allows examining the transition effects in more detail. For good statistics, one should collect each spectrum for a longer period of time or one should have a more intense source, but very interesting results can be obtained in a reasonable time even at a first generation Spallation source like IPNS.

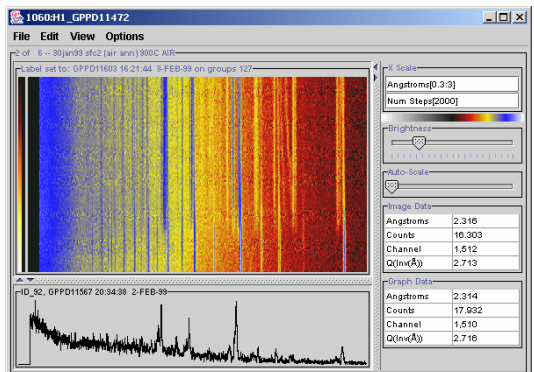

Fig. 3: An image view of a data set created by merging spectra from 120 run files. The streaks clearly show phase changes as a function of the Oxygen partial pressure.

## V. Texture studies on diffractometers

As mentioned earlier, at a time-of-flight diffractometer, one can usually focus a bank of detectors to match a reference scattering angle, either by proper placement of detectors, or by scaling the measured time. This will allow the diffraction peaks of a large bank of detectors to all line up at the same position when the x-axis is converted to lattice spacing (d-value) or wave vector (Q). If the sample is a powder with randomly oriented crystallites, the vertical lines for a given diffraction peak will have fairly uniform intensity from one detector element to another. However, if there is texture in the sample so the scattering is non-isotropic, the intensity of a given peak will vary considerably with the detector scattering angle. Keeping each detector separate allows studying sample texture. A diffraction pattern of a sample (YBCO) with a great deal of texture is shown in Fig. 4.

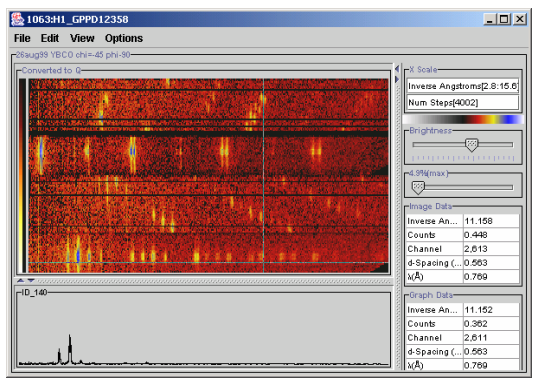

Figure 4. An image view of a sample of YBCO which has a great deal of texture shown by the anisotropy in the scattering intensity. Without anisotropy the vertical lines would traverse the entire image.

## VI. Single crystal time-of-flight data analysis

Single crystal time-of-flight neutron scattering data cannot be reduced by focusing in the same way that powder diffraction data is focused. Reduction of single crystal data requires actually finding the peaks in the data set and saving information about the peaks instead of saving the raw data. The first time-of-flight single crystal diffractometer was developed at IPNS and there is a complete set of programs for finding the peaks in the data, fitting the peaks, indexing them, creating an orientation matrix for the crystal, etc (A. J. Schultz, et al., 1984). These programs were initially written in VAX Fortran and have been subsequently converted to GNU Fortran on Linux. When we surveyed the existing software, we decided that two of the SCD analysis programs would need to be rewritten in the Java language since they needed to read the raw data files and we did not want to rewrite the routines for reading the data. The rest of the programs could initially be run from ISAW by wrapping the call to the program with a Java routine to get the parameters needed by the Fortran routine.

The system interface provided by the Java language and available through an ISAW operator, allows executing any program, whether created by Fortran or some other method. Although the area detector covers a fairly large region of space, it currently takes measurements at 44 different crystal orientations to cover most of reciprocal space. Scripted routines for reading all the data and rotating it to a common orientation will allow rapid data analysis. We are already through the steps of reading the data, finding the peaks, and indexing the peaks. The rest of the SCD data analysis is expected to be integrated with ISAW by early next year.

The image view discussed previously is useful for finding where peaks occur in time or d-spacing if the area detector pixels are considered as a set of time-of-flight spectra, but we also have a raster/contour viewer (Fig. 5) which can show the response of the detector at a given time. Coordination of the image view and raster view through the cursor position simplifies finding of interesting features. Pointing at a peak in the image view causes an open raster view to shift to the correct time slice to view the peak.

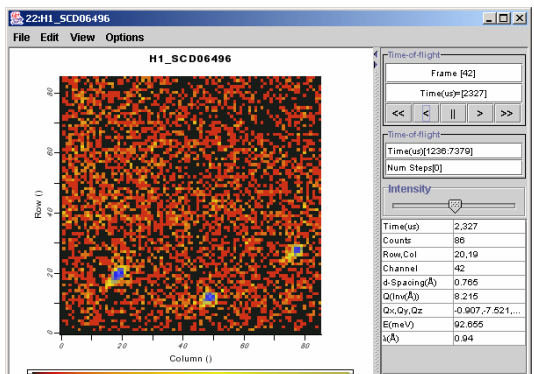

Fig. 5: Raster view of an SCD data set. Rows and columns of the raster image correspond to the rows and columns of the pixels of the area detectors. The forward and backward controls allow stepping through time-slices.

Because of the large quantity of data contained in a single crystal diffractometer data set, one of us (D. Mikkelson) developed a special viewer that would efficiently handle the sparse data in real time by rotating each set of measurements to the same coordinate system. We are in the process of integrating that special viewer with ISAW. The new viewer is similar to the 3D viewer in that it allows you to rotate the data in real time.

## VII. Conclusions

ISAW satisfies all of our requirements for a software package for reducing and visualizing data from next generation neutron scattering instruments. It runs on most computer platforms, is freely available, easy to use, and includes network access to data servers. It provides viewers and operations for handling thousands of spectra and for adding and/or merging data from multiple files. It can write data in NeXus, XML, or ASCII formats.

On today's average personal computer with 256 MB of memory, ISAW can handle data sets with up to 10000 spectra. Current "state of the art" small servers can have 10-20 times this amount of memory. If the speed and capacity of computers continues to double every one and a half years, we could expect to handle million pixel data sets in 4-5 years. To handle larger data sets or to handle this size of data set with today's computer, the data needs to be divided into chunks. We have already found 256-512 MB of memory insufficient for some of the data sets from LANSCE and are in the process of modifying ISAW so we can read portions of the data instead of all the data at once. The LANSCE data contains data sets for pulse height spectra, beam monitors, and scattering data. As a first step to reading selected parts of data files, we are modifying ISAW data retrievers to present the option of loading only selected data sets. We will also allow loading selected spectra within those data sets. This will save memory by only loading data that we are interested in. We will also be able to avoid loading data not of interest such as data from dead detectors . These developments are also important for the live data servers to reduce the amount of data that must be sent over the network.

## VIII. Acknowledgements

Work performed at ANL is supported by the U.S. DOE-BES under contracts No. W-31-109-ENG-38. Partial support was also provided by the Division of Educational programs of Argonne National Laboratory, and by the National Science Foundation, award number DMR-0218882.